\documentclass{article}
\usepackage{lnfprep}
\usepackage{graphicx}
\usepackage{subfigure}
\usepackage{amsmath}
\usepackage{amssymb}

\def\beq{\begin{equation}}   \def\eeq{\end{equation}}
\def\bea{\begin{eqnarray}}   \def\eea{\end{eqnarray}}

\newcommand{\gsim}{\lower.7ex\hbox{$ \;\stackrel{\textstyle>}{\sim}\;$}}
\newcommand{\lsim}{\lower.7ex\hbox{$ \;\stackrel{\textstyle<}{\sim}\;$}}

\def\c2{CLEO~II.V}

\def\d0d0{ D^0\bar{D}^0 }
\def\p0p0{ P^0\bar{P}^0 }

\def\qp2{ \Bigl| \frac{q}{p} \Bigr|^2 }
\def\pq2{ \Bigl| \frac{p}{q} \Bigr|^2 }

\def\ps2s{  \psi(2S) }
\def\q2{ $q^2$ }
\def\cm2s1{ $\,{\rm cm}^{-2} {\rm s}^{-1}$}
\def\d0{D_2^{*0}}
\def\d+{D_2^{*+}}

\newcommand{\Header}{
  \begin{tabular}{rl}
  \hspace{-.4cm}
      &
    \renewcommand{\arraystretch}{0.5}
    \renewcommand{\arraystretch}{1}
  \end{tabular}
  \vskip 1cm
  \begin{flushright}
  \renewcommand{\arraystretch}{0.5}
    \begin{tabular}{r}
      {\underline{LNF-05/30 (P)}}\\    
      {\small 22 dicembre 2005} \\      
    \end{tabular}
  \end{flushright}
  \renewcommand{\arraystretch}{1}
  \vskip 1 cm
  }

\begin{document}
\begin{titlepage}
\title{
  \Header
  {\LARGE  \textsc{\textmd {MICROMETRIC POSITION MONITORING USING FIBER BRAGG GRATING
  SENSORS IN SILICON DETECTORS}}
  }
}
\author{E.Basile(*), F.Bellucci (***), L.
Benussi, M. Bertani, S. Bianco, M.A. Caponero (**),  \\
D. Colonna (*), F. Di Falco (*), F.L. Fabbri, F. Felli (*), M.
Giardoni, A. La Monaca, \\
F.Massa (*), G. Mensitieri (***), B. Ortenzi, M. Pallotta, A.
Paolozzi (*), L.
Passamonti, \\
D. Pierluigi, C. Pucci (*), A. Russo, G. Saviano (*)\dag.\\
{\em Laboratori Nazionali di Frascati dell'INFN, v.E.Fermi 40 00044
Frascati (Rome) Italy } \\
~ \\
presented by S.Bianco at ICATPP05, Villa Olmo (Como) Italy 2005
} \maketitle \baselineskip=1pt

\begin{abstract}
\indent We show R\&D results including long term stability,
resolution, radiation hardness and characterization of Fiber Grating
sensors used to monitor structure deformation, repositioning and
surveying of silicon detector in High Energy Physics.
\end{abstract}

\vspace*{\stretch{2}}

\vskip 1cm
\begin{flushleft}
\begin{tabular}{l l}
  \hline\\
  $ ^{*\,\,\,\,\,\,}$ & \footnotesize{Permanent address: ``La Sapienza" University - Rome.} \\
  $ ^{**\,\,\,}$& \footnotesize{Permanent address: ENEA Frascati.} \\
  $ ^{***}$ & \footnotesize{Permanent address: ``Federico II" University - Naples.} \\
\end{tabular}
\end{flushleft}
\dag This work was supported by the Italian Istituto Nazionale di
Fisica Nucleare and Ministero dell'Istruzione, dell'Università e
della Ricerca. This work was partially funded by contract EU
RII3-CT-2004-506078.
\end{titlepage}
\pagestyle{plain}
\setcounter{page}2
\baselineskip=17pt

\section{  \textsc{Introduction}}
FBG sensors are widely used in telecommunication as optical filters.
For the first time we have used FBG sensors as optical, low-noise,
high-resolution strain gauges to monitor structure deformation,
repositioning, and surveying silicon (pixel and microstrips)
detectors for HEP experiments at hadron machines. We show R\&D
results including long term stability, precision,
resolution, radiation hardness and characterization.\\
\newline
\section{  \textsc{FBG Sensors}}
Fiber Bragg Grating (FBG) sensors have been used so far as
telecommunication filters, and as optical strain gauges in civil and
aerospace engineering [1], and, only recently, in HEP detectors [2].
The BTeV [3] detectors utilize FBG sensors to monitor online the
position of the straw tubes, pixels, and microstrip. The optical
fiber is used for monitoring displacements and strains in mechanical
structures such as the straw tube-microstrip support presented here.
A modulated refractive index along the FBG sensor produces Bragg
reflection at a wavelength dependent on the strain in the fiber
(Fig.1), permitting real-time monitoring of the support. According
to these properties, an FBG sensor is going to be plced in the M0X
structure between the Rohacell$\circledR$ foam and the CFRP shell.
Sensors will be located in spots of maximal deformation, as
predicted by FEA simulation. Figure 2 shows long-term behaviour of
FBG sensors while monitoring micron-size displacements, compared to
monitoring via microphotographic methods.
 \newline
\section{  \textsc{Long-term Stability and Radiation Hardness}}
The optical fiber is used for monitoring displacements and strains
in mechanical structures such s the presented straw tubes-microstrip
support. A wavelength selective light diffraction grating (Fig.1)
along the FBG sensor is placed in the fiber, and it permits an
on-time monitoring of the support. Fig.2 shows long-term behaviour
of FBG sensors while monitoring micron-size desplacements, compared
to monitoring via photografic methods.\\
Sensors have been tested for radiation damage. Fig.3 shows spectral
response up to a neutron fluence of $1.6\cdot10^{13}$ 14-MeV
neutrons/cm$^{2}$, corresponding to 6 months BTeV integrated dose.
\newline
\section{  \textsc{The Omega-like Repositioning Device}} FBG
sensors have been also applied to instrument a novel repositioning
device with micrometric resolution. The Omega-like device (shown as
prototype in Fig.4,5) follows the displacement of the pixel detector
designed for the BTeV experiment at the Fermila Tevatron which, at
each accelerator store, has to be moved out and in of the beamline.
Fig.6 shows a Finite Element Analysis of the Omega-like device. FBG
sensor are located on area of largest strain in order to maximize
sensitivity. Preliminary results show how a repositioning precision
of about $10\,\mu$m  is reached. Work is in progress to reach the
required $3\, \mu$m  precision.
 \newline
\begin{figure}[!htbp]
\begin{center}
  \includegraphics[width=7cm]{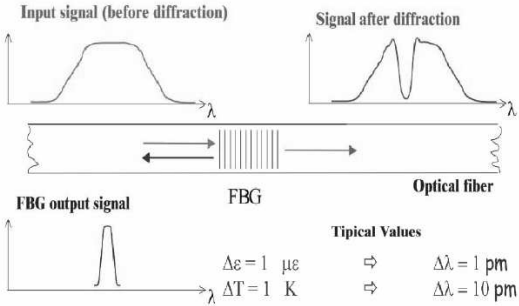}\\
  \caption{Principle of FBG sensors operation. A laser pulse is injected in the fiber
  and reflected selectively according to the grating pitch. Strain $\Delta\varepsilon$ changes
  the grating pitch thus changing the wavelength of reflected pulse. The sensors is also
  sensitive to temperature changes.}
  \end{center}
\end{figure}

\

\begin{figure}[!htbp]
\begin{center}
  \includegraphics[width=7cm]{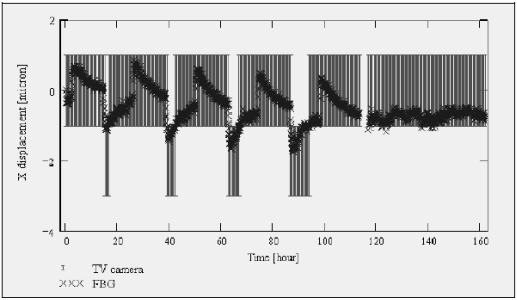}\\
  \caption{FBG long-term monitoring stability results. FBG output (crosses) is validated by TV camera (bars).}
  \end{center}
\end{figure}

\begin{figure}
\begin{center}
  \includegraphics[width=7cm]{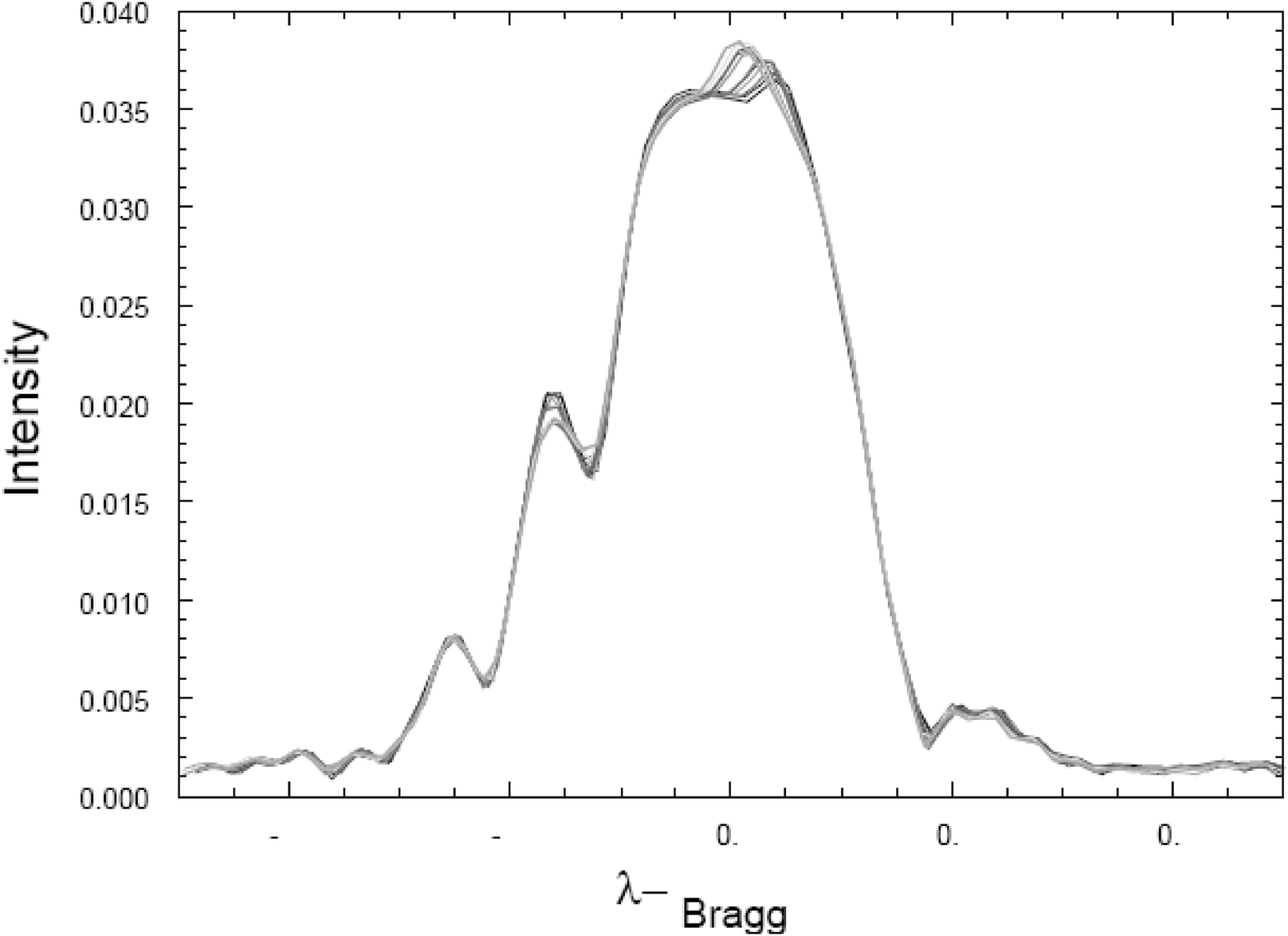}\\
  \caption{Radiation hardness of FBG sensors. Spectral response up to neutron fluence of $1.6\cdot10^{13}$ 14-MeV
neutrons/cm$^{2}$, corresponding to 6 months BTeV integrated dose. No frequency
shift is observed up to max irradiated dose.}
  \end{center}
\end{figure}

\begin{figure}
\begin{center}
  \includegraphics[width=7cm]{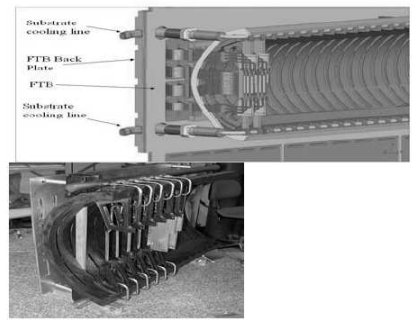}\\
  \caption{Sketch of BTeV pixel detector and its Carbon Fiber Reinforced Plastic support frame.}
  \end{center}
\end{figure}

\begin{figure}
\begin{center}
  \includegraphics[width=5cm]{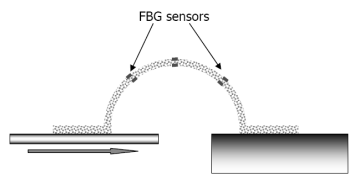}\\
  \caption{The Omega-like repositioning device, equipped with FBG sensors, follows the pixel support structure
  in and out of beam, assuring repositioning accuracy.}
  \end{center}
\end{figure}

\begin{figure}
\begin{center}
  \includegraphics[width=5cm]{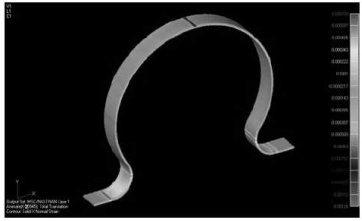}\\
  \caption{Finite Element Analysis of the Omega-like repositioning system. Sensors are located on the
  areas of larger mechanical strain in order to maximize sensitivity.}
  \end{center}
\end{figure}

\newpage
\section{  \textsc{Conclusions}}
We have used FBG sensor in HEP for the first time as precise,
stable, optical devices for micrometric position monitoring of
silicon pixel and strips detectors. FBG sensors provide position
monitoring with micrometric resolution. Under radiation with doses
typical of year-long operation at hadron colliders they show no sign
of spectral response shift. We have used sensors to characterize and
optimize pixel support structures in Carbon Fiber Reinforced
Plastic. Finally, we have proposed a novel device to precisely
reposition the pixel detector in and out of the beams at each
accelerator store. Preliminary results show a $10\mu$m resolution,
improvements are undergoing and we expect to  reach the $3\mu$m
precision required by the experimental operation.
\newline
%

\newpage

\end{document}